\documentclass[letterpaper,11pt]{article}

\usepackage{amssymb}
\usepackage{cite}

\newcommand {\beq} {\begin{equation}}
\newcommand {\eeq} {\end{equation}}
\newcommand {\beqa} {\begin{eqnarray}}
\newcommand {\eeqa} {\end{eqnarray}}
\newcommand {\beqan} {\begin{eqnarray*}}
\newcommand {\eeqan} {\end{eqnarray*}}

\newcommand {\nn} {\nonumber}
\newcommand {\ph}[1]{\phantom{#1}}
\newcommand {\ie}{i.e.~}

\newcommand {\sss} {\scriptscriptstyle}

\newcommand{\ordo}{\ensuremath{\mathcal{O}}}

\newcommand{\al}{\ensuremath{\alpha}}
\newcommand{\be}{\ensuremath{\beta}}
\newcommand{\ga}{\ensuremath{\gamma}}
\newcommand{\Ga}{\ensuremath{\Gamma}}
\newcommand{\de}{\ensuremath{\delta}}

\newcommand{\eps}{\ensuremath{\epsilon}}
\newcommand{\veps}{\ensuremath{\varepsilon}}

\newcommand{\si}{\ensuremath{\sigma}}
\newcommand{\Si}{\ensuremath{\Sigma}}
\newcommand{\om}{\ensuremath{\omega}}
\newcommand{\Om}{\ensuremath{\Omega}}

\newcommand{\mathHb}[1]{{\mathop{\kern0pt#1}\limits^{\,\sss
      \prime\prime}\vphantom{#1}}}

\newcommand {\we} {\wedge}
\newcommand {\pa} {\partial}

\newcommand{\eqnlab}[1]{\label{eqn:#1}}

\newcommand{\eqnref}[1]{(\ref{eqn:#1})}

\newcommand{\Eqnref}[1]{Eq.~(\ref{eqn:#1})}

\newcommand{\Eqsref}[1]{Eqs.~(\ref{eqn:#1})}

\newcommand{\seclab}[1]{\label{sec:#1}}

\newcommand{\Secref}[1]{Section~\ref{sec:#1}}

\begin{document}

 \pagestyle{empty}
 \vskip-10pt
 \hfill {\tt hep-th/0309244}

\begin{center}

\vspace*{2cm}

% \rule{150mm}{2pt}
%\vskip 0.5truecm
\noindent
{\LARGE\textsf{\textbf{Supersymmetric coupling of a self-dual string
      \\[5mm] to a $(2, 0)$ tensor multiplet background}}}
%\vskip 2truecm
%Version: \today
\vskip 1truecm
% \rule{150mm}{2pt}
\vskip 2truecm

{\large \textsf{\textbf{P\"ar Arvidsson\footnote{\tt par@fy.chalmers.se}, Erik
  Flink\footnote{\tt erik.flink@fy.chalmers.se} and M{\aa}ns
  Henningson\footnote{\tt mans@fy.chalmers.se}}}} \\ 
\vskip 1truecm
{\it Department of Theoretical  Physics\\ Chalmers University of
  Technology and G\"oteborg University\\ SE-412 96 G\"{o}teborg,
  Sweden}\\
\end{center}
\vskip 1cm
\noindent{\bf Abstract:}
We construct an interaction between a $(2, 0)$ tensor multiplet in six
dimensions and a self-dual string. The interaction is a sum of a
Nambu-Goto term, with the tension of the string given by the modulus
of the scalar fields of the tensor multiplet, and a non-local
Wess-Zumino term,
that encodes the electromagnetic coupling of the string to the two-form
gauge field of the tensor multiplet. The interaction is invariant under
global $(2, 0)$ supersymmetry, modulo the equations of motion of a free 
tensor multiplet. It is also invariant under a local fermionic
$\kappa$-symmetry, as required by the BPS-property of the string.
\newpage
\pagestyle{plain}

\section{Introduction} 
The six-dimensional $(2, 0)$ theories are one of the most mysterious
discoveries in string theory during the past decade, and finding the
right framework to define them remains an outstanding challenge
\cite{Witten:1995}. In previous publications~\cite{AFH:2002,AFH:2003},
we have pursued an approach
to study these theories by working above a generic vacuum state, \ie
not at the origin of the moduli space, where they can be thought of as
describing the dynamics of massless particles and tensile strings. 
The existence of these constituents follows directly from the $(2, 0)$ 
supersymmetry algebra \cite{Gustavsson:2001sr}. At energies low
compared to the scale set by the string tension, the theory is free and 
well understood, but at higher energies the interactions between the
particles and the strings must be taken into account. In the present paper, 
we will construct the leading such interactions.

To be more specific, we will focus our attention on the simplest
$A_1$ version of $(2, 0)$ theory. The particles then arise upon
quantization of a single $(2, 0)$ tensor multiplet, \ie a set of scalar
fields $\phi$ in the ${\bf 5}$ representation of the SO(5)
$R$-symmetry group, a set of chiral spinors $\psi$ in the ${\bf 4}$
representation of SO(5), and a two-form gauge field $b$ in the
singlet ${\bf 1}$ representation of SO(5). The field strength $h =
d b$ of $b$ can be decomposed into its self-dual part $h_+ =
\frac{1}{2} (h + * h)$ and anti self-dual part $h_- = \frac{1}{2} (h -
* h)$. The anti self-dual part $h_-$ is actually not part of the
tensor multiplet, but it is convenient to include it nevertheless as a
decoupled `spectator' field. One can then construct a supersymmetric
free action for these fields  \cite{AFH:2003}:
\beq
S_{TM} = \int_M \left( d \phi \cdot \wedge * d \phi + h \wedge * h +
\psi \; \slash \!\!\!\!  D \psi \right), \eqnlab{TM}
\eeq
where the integral is over six-dimensional Minkowski space $M$ and the
bilinears in the fields are SO(5) invariants. The coefficients of
the kinetic terms for $\phi$ and $\psi$ can be absorbed by changing
the normalization of these fields (which take their values in linear
spaces) and are thus of no consequence. However, $b$ is not really a
tensor field but rather a `connection', so its normalization has an
absolute meaning. The coefficient of its kinetic term is a priori a
parameter of the theory, but it turns out that its value is uniquely
determined by the consistent decoupling of $h_-$
\cite{Henningson:2002qh}.

The $A_1$ version of $(2, 0)$ theory also contains a single type of
string. The degrees of freedom of such a string is an embedding map
$X$ from the string world-sheet $\Sigma$ to Minkowski space $M$ and a
set of world-sheet fermions $\Theta$ that transform as an anti-chiral
spinor in Minkowski space and in the ${\bf 4}$ of SO(5). Actually,
two of the six components of $X$ can be eliminated by fixing the
reparametrization invariance of the world-sheet $\Sigma$. Similarly,
half of the components of $\Theta$ can be eliminated by fixing a local
fermionic $\kappa$-symmetry on $\Sigma$.

The couplings between the world-sheet degrees of freedom $X$ and
$\Theta$ and the Minkowski space fields $\phi$, $b$, and $\psi$ are
dictated by the following three principles~\cite{AFH:2002}:
\begin{itemize}
\item
The tension of the string should be given by the local value of
$\sqrt{\phi \cdot \phi}$.
\item
The string should couple electrically and magnetically to $b$ in such
a way that the anti self-dual part $h_-$ of $h = d b$ remains
decoupled.
\item
The interactions should be supersymmetric.
\end{itemize}
In the present paper, we will not solve this problem completely.
What we will do is to construct an interaction that is supersymmetric
provided that the tensor multiplet fields fulfill the free equations
of motion following from the action \eqnref{TM}. But when this interaction
is added to the action, the equations of motion are altered and the
interaction is no longer supersymmetric. Nevertheless, our model
can be trusted for computing scattering of tensor multiplet particles off 
a string to lowest order in their energy, and we intend to present
such a calculation in a forthcoming publication. We also believe that
the present work is a step towards constructing the exactly supersymmetric 
model, and hope to be able to return to this issue in the future. 

To motivate our construction, it is convenient to first neglect the 
fermionic degrees of freedom, \ie $\psi$ and $\Theta$, and concentrate on 
fulfilling the first two requirements. 
It is then straightforward to comply with the first requirement by
constructing a kinetic term for the string of Nambu-Goto type:
\beq
S_{\sss NG} = \int_\Sigma d^2 \sigma \sqrt{\phi \cdot \phi} \sqrt{-G}
. \eqnlab{NG}
\eeq 
Here, the world-sheet $\Sigma$ is parametrized by $\sigma$, and $G$ is
the determinant of
the metric on $\Sigma$ induced by its embedding into Minkowski space
$M$. The coefficient of this term will eventually be determined by
the requirement of $\kappa$-symmetry.

To incorporate the electromagnetic coupling, we start with the
standard electric coupling of the string to the field $b$:
\beq
S_{el} = \int_\Sigma b ,
\eeq
where a pull-back of the Minkowski space field $b$ to the world-sheet 
$\Sigma$ by the embedding map is understood. Let now
$D$ be an open three-manifold embedded in Minkowski space $M$ such
that its boundary is given by $\Sigma$, \ie $\partial D = \Sigma$. We
can then rewrite the electric coupling by using Stokes' theorem as
\beq
S_{el} = \int_D h ,
\eeq
where a pull-back to $D$ is understood. This expression for the
electric coupling can be used also in a topologically non-trivial
context, where the connection $b$ cannot be represented by a globally
defined two-form. One must then ensure that, given $\Sigma$, the
choice of three-manifold $D$ has no observable consequences. In the
usual way, this means that the coefficient of the electric coupling is
quantized to integer values. Given this form of the electric coupling, 
it is natural to introduce a corresponding magnetic coupling
as
\beq
S_{mag} = \int_D * h .
\eeq
The three-manifold $D$ can thus be interpreted as the world-volume of
a 'Dirac membrane' ending on the string~\cite{Deser:1998}.
Altogether, we have an electromagnetic coupling of Wess-Zumino type:
\beq
S_{\sss WZ} =  \int_D f , \eqnlab{WZ}
\eeq
where $f = (h + * h) = 2 h_+$. This term thus manifestly leaves 
the anti self-dual part $h_-$ of $h$ decoupled as required. It should
be noted that the three-form $f$ is closed, $d f = 0$, when the
field strength $h$ fulfills the Bianchi identity $d h = 0$ and the
free equation of motion $d (*h) = 0$.

The supersymmetrization of the interaction terms $S_{\sss NG}$ and
$S_{\sss WZ}$ 
is analogous to many similar situations where branes are coupled
to background fields
\cite{Bergshoeff:1987,Cederwall:1997_3,Cederwall:1997_p}.
Following~\cite{Howe:1983},
we introduce $(2, 0)$ superspace
and a constrained superfield $\Phi$ with component fields 
$\phi$, $\psi$ and $h_+$. We also construct
a closed super-three-form $F$ out of $\Phi$. 
The string degrees of freedom $X$ and
$\Theta$ are considered as describing the embedding of the string
world-sheet $\Sigma$ into superspace. Also, $\Sigma$ is the
boundary of a three-manifold $D$ embedded into superspace. In this
way, the interaction terms can still be written as the sum of a
Nambu-Goto term \eqnref{NG} and a Wess-Zumino term \eqnref{WZ},
with $\phi$ and $f$ replaced by the superfields $\Phi$ and $F$.
Supersymmetry is then manifest for each of these terms separately.
The linear combination $S_{\sss NG} + S_{\sss WZ}$ is also invariant
under a
local fermionic $\kappa$-symmetry. The parameter of this symmetry
has the same representation content as $\Theta$, but a certain projection
has to be imposed that eliminates half of the components. The remaining
symmetry is thus precisely sufficient to put half of the components
of $\Theta$ to zero, as required by supersymmetry.

\section{The supergeometry}

\subsection{Superspace}
\seclab{superspace}

Following~\cite{Howe:1983}, we introduce a superspace with bosonic
  coordinates $x^{\al 
  \be}=-x^{\be \al}$ and fermionic coordinates $\theta^{\al}_a$, where
  $\al,\be=1,\ldots,4$ are SO(5,1) chiral spinor indices and
  $a=1,\ldots,4$ is an SO(5) spinor index. The latter index is
  raised and lowered from the left using the antisymmetric SO(5)
  invariant tensor $\Om^{ab}$ and its inverse $\Om_{ab}$, \ie
\beqa
\theta^{a \al} & = & \Om^{ab} \theta^{\al}_b \\
\theta^{\al}_a & = & \Om_{ab} \theta^{b \al}.
\eeqa
For consistency, this requires that $\Om_{ab} \Om^{bc} =
  \de_{a}^{\ph{a} c}$.

Supersymmetry transformations are generated by the supercharges
\beq
\eqnlab{supercharge}
Q^a_{\al} = \pa^a_{\al} - i \Om^{ab} \theta^{\be}_b \pa_{\al \be},
\eeq
where $\pa_{\al \be}=-\pa_{\be \al}$ denotes the ordinary bosonic
derivative and $\pa^a_{\al}$ a derivative with respect to the
fermionic coordinate $\theta_a^{\al}$. The supercharges obey the
anticommutation relations
\beq
\left\{ Q^a_{\al} , Q^b_{\be} \right\} = -2i \Om^{ab} \pa_{\al \be}.
\eeq
We also need the superspace covariant derivatives
\beq
\eqnlab{superd}
D^a_{\al} = \pa^a_{\al} + i \Om^{ab} \theta^{\be}_b \pa_{\al \be},
\eeq
obeying
\beq
\left\{ D^a_{\al} , D^b_{\be} \right\} = 2i \Om^{ab} \pa_{\al \be}.
\eeq

\subsection{Superfields}
\seclab{superfields}
 
After these preliminaries, we introduce the superfield
$\Phi^{ab}=-\Phi^{ba}$ which is a function of the superspace
coordinates $x$ and $\theta$. This superfield is subject to the
algebraic constraint 
\beq
\eqnlab{algcon}
\Om_{ab} \Phi^{ab}=0,
\eeq 
and the differential constraint
\beq
\eqnlab{diffcon}
D^a_{\al} \Phi^{bc} + \frac{1}{5} \Om_{de} D^d_{\al} \left( 2 \Om^{ab}
\Phi^{ec} - 2 \Om^{ac} \Phi^{eb} + \Om^{bc} \Phi^{ea} \right) = 0.
\eeq

It is useful to introduce two additional superfields $\Psi^a_{\al}$
and $H_{\al \be}$, related to $\Phi^{ab}$ by the relations
\beqa
\Psi^c_{\al} & = & - \frac{2i}{5} \Om_{ab} D_{\al}^a \Phi^{bc} \\
H_{\al \be} & = & \frac{1}{4} \Om_{ab} D^a_{\al} \Psi^b_{\be}.
\eeqa
It can be shown that $H_{\al \be} = H_{\be \al}$. It thus transforms
as a self-dual three-form. (An anti self-dual three-form would be
denoted by $H^{\al \be}$.) The differential
constraint \eqnref{diffcon} yields that
\beqa
\eqnlab{diffcon_Phi}
D^a_{\al} \Phi^{bc} & = & - i \left(\Om^{ab} \Psi^c_{\al} +
\Om^{ca} \Psi^b_{\al} + \frac{1}{2} \Om^{bc} \Psi^a_{\al} \right) \\
\eqnlab{diffcon_Psi}
D^a_{\al} \Psi^b_{\be} & = & 2 \pa_{\al \be} \Phi^{ab} - \Om^{ab}
H_{\al \be} \\
\eqnlab{diffcon_H}
D^a_{\al} H_{\be \ga} & = & i \left( \pa_{\al \be} \Psi^a_{\ga} +
\pa_{\al \ga} \Psi^a_{\be} \right),
\eeqa
along with the equations of motion
\beqa
\eqnlab{eom_Phi}
\pa_{[\al \be} \pa_{\ga \de]} \Phi^{ab} & = & 0 \\
\eqnlab{eom_Psi}
\pa_{[\al \be} \Psi^a_{\ga]} & = & 0 \\
\eqnlab{eom_H}
\pa_{[\al \be} H_{\ga] \de} & = & 0.
\eeqa
The fact that the equations of motion appear is a consequence of
having on-shell supersymmetry (\ie the supersymmetry algebra only
closes on-shell).

Denoting the lowest components (\ie those independent of
$\theta$) of $\Phi$, $\Psi$ and $H$ by $\phi$, $\psi$ and $h$,
respectively, we can obtain an explicit expression for $\Phi$ in terms
of these:
\beqa
\Phi^{ab} & = & \phi^{ab} -i \theta^{\al}_c \Big(\Om^{ca}
\psi^b_{\al} + \Om^{bc} \psi^a_{\al} + \frac{1}{2}\Om^{ab}
\psi^c_{\al} \Big)+{} \nn \\
&&{}+ i  \theta^{\al}_c \theta^{\be}_d \Big( h_{\al \be}
\big( \Om^{da}\Om^{bc} + \frac{1}{4}\Om^{dc}\Om^{ab} \big) - \Om^{da}
\pa_{\al\be} \phi^{bc} - \Om^{db} \pa_{\al\be} \phi^{ca} \Big) + {}
\nn \\
& & {} + \frac{1}{6} \theta^{\al}_c \theta^{\be}_d \theta^{\ga}_e
\Big( \Om^{ab} \Om^{ce} \pa_{\al \be} \psi^d_{\ga} + \Om^{ac} \big(
\Om^{bd} \pa_{\al \be} \psi^e_{\ga} + 2  \Om^{de} \pa_{\al \be}
\psi^b_{\ga} + 4 \Om^{eb} \pa_{\al \be} \psi^d_{\ga} \big) - {} \nn \\
& & {} - \Om^{bc} \big( \Om^{ad} \pa_{\al \be} \psi^e_{\ga} + 2
\Om^{de} \pa_{\al \be} \psi^a_{\ga} + 4 \Om^{ea} \pa_{\al \be}
\psi^d_{\ga} \big) \Big) + \ordo(\theta^4),
\eeqa
where all higher order terms can be expressed in terms of $\phi$,
$\psi$ and $h$. A priori, the final term in this sum is of order
$\theta^{16}$.

A general superfield transforms according to
\beq
\de \Phi^{ab} = [\eta_c^{\ga} Q_{\ga}^c , \Phi^{ab}]
\eeq
under a supersymmetry variation with a constant fermionic parameter
$\eta$. For the lowest components of $\Phi$, $\Psi$ and $H$ we get that
\beqa
\eqnlab{susy_phi}
\de \phi^{ab} & = & i \left( \Om^{ac} \eta^{\al}_c \psi^b_{\al} -
\Om^{bc} \eta^{\al}_c \psi^a_{\al} - \frac{1}{2} \eta_c^{\al}
\psi^c_{\al} \Om^{ab} \right) \\
\eqnlab{susy_psi}
\de \psi^a_{\al} & = & \Om^{ab} \eta_b^{\be} h_{\al \be} + 2 \pa_{\al
  \be} \phi^{ab} \eta_b^{\be} \\
\eqnlab{susy_h}
\de h_{\al \be} & = & - i \eta_a^{\ga} \left( \pa_{\al \ga}
\psi^a_{\be} + \pa_{\be \ga} \psi^a_{\al} \right),
\eeqa
which coincides with the supersymmetry transformations of the tensor
multiplet fields obtained in~\cite{AFH:2003}.

The free action
\beq
S_{\sss TM} = \int d^6 x \left\{ - \Om_{ac} \Om_{bd} \pa_{\al \be}
\phi^{ab} \pa^{\al \be} \phi^{cd} + 2 h_{\al \be} h^{\al \be} - 4i
\Om_{ab} \psi^a_{\al} \pa^{\al \be} \psi^b_{\be} \right\}
\eeq
is manifestly SO(5) invariant, and one may check that it is also
invariant under the above supersymmetry transformations. The equations
of motion following from this action coincide with (the lowest
components of) \Eqsref{eom_Phi}-\eqnref{eom_H}. For a more elaborate
discussion on these issues, along with reality and SO(5)-properties of
the fields, we refer to~\cite{AFH:2003}.

\subsection{Superforms}

The next step is to set the stage for superforms. Let $z^{\sss{M}}$ be
a (curved) coordinate in superspace, where ${}^{\sss{M}}$ may take
values ${}^{[\mu \nu]}$ (bosonic) and ${}^{\mu}_{m}$
(fermionic). Explicitly, we have the differentials
\beqa
dz^{[\mu \nu]} & = & dx^{\mu \nu} \\
dz^{\mu}_{m} & = & d \theta^{\mu}_{m}.
\eeqa

We also introduce the tangent space differentials $E^{\sss{A}}$
related to $dz^{\sss{M}}$ by
\beq
E^{\sss{A}} = dz^{\sss{M}} E_{\sss{M}}^{\ph{\sss{M}}\sss{A}} \qquad
\Leftrightarrow \qquad dz^{\sss{M}} = E^{\sss{A}}
E_{\sss{A}}^{\ph{\sss{A}}\sss{M}},
\eeq
where $E_{\sss{M}}^{\ph{\sss{M}}\sss{A}}$ is the vielbein and
its inverse is denoted by $E_{\sss{A}}^{\ph{\sss{A}}\sss{M}}$.
The tangent space index~${}^{\sss{A}}$ also takes bosonic
(${}^{[\al \be]}$) and fermionic (${}^{\al}_{a}$)
values. (The coordinates $z^{\al \be}=x^{\al \be}$ and
$z^{\al}_a=\theta^{\al}_a$ are the same as those introduced in
\Secref{superspace}.) In our case, we can choose the tangent space
  differential superforms as
\beqa
\eqnlab{P_def}
E^{[\al \be]} & = & dx^{\al \be} + i \Om^{ab} \theta^{[\al}_a d
  \theta^{\be]}_b \\
E^{\al}_a & = & d \theta^{\al}_a,
\eeqa
meaning that the vielbein $E_{\sss{M}}^{\ph{\sss{M}}\sss{A}}$ has the
components
\beqa
E_{[\mu \nu]}^{\ph{[\mu \nu]} [\al \be]} & = & \de_{\mu}^{\ph{\mu}
  [\al} \de_{\nu}^{\ph{\nu} \be]} \\
E_{[\mu \nu] a}^{\ph{[\mu \nu]} \al} & = & 0 \\
E_{\mu}^{m [\al \be]} & = & - i \de_{\mu}^{\ph{\mu} [ \al}
  \de_{\nu}^{\ph{\nu} \be ]} \Om^{mn}
  \theta_n^{\nu} \\
E_{\mu a}^{m \al} & = & \de_{\mu}^{\ph{\mu} \al} \de^{m}_{\ph{m}a},
\eeqa
while the inverse vielbein $E_{\sss{A}}^{\ph{\sss{A}}\sss{M}}$ is
\beqa
E_{[\al \be]}^{\ph{[\al \be]} [\mu \nu]} & = & \de_{\al}^{\ph{\al}
  [\mu} \de_{\be}^{\ph{\be} \nu]} \\
E_{[\al \be] m}^{\ph{[\al \be]} \mu} & = & 0 \\
E_{\al}^{a [\mu \nu]} & = & i \de_{\al}^{\ph{\al} [ \mu}
  \de_{\be}^{\ph{\be} \nu]} \Om^{ab}
  \theta_b^{\be} \\
E_{\al m}^{a \mu} & = & \de_{\al}^{\ph{\al} \mu} \de^{a}_{\ph{a}m}.
\eeqa
The latter set of equations yields that the derivatives in tangent
space coordinates,
\beq
D_{\sss{A}} \equiv E_{\sss{A}}^{\ph{\sss{A}}\sss{M}} \pa_{\sss{M}},
\eeq
become
\beqa
D_{[\al \be]} & = & \pa_{\al \be} \\
D^a_{\al} & = & \pa^a_{\al} + i \Om^{ab} \theta^{\be}_b \pa_{\al \be}.
\eeqa
The fermionic part agrees with the superderivative in
\Eqnref{superd}.

A super-$n$-form is now written as
\beq
\om = \frac{1}{n!} dz^{{\sss M_n}} \we \ldots \we dz^{{\sss M_1}}
\om_{{\sss M_1 \ldots M_n}},
\eeq
where the wedge product is such that $dz^{\sss M} \we dz^{\sss N} =
dz^{\sss N} \we dz^{\sss M}$ if both $M$ and $N$ are fermionic,
otherwise $dz^{\sss M} \we dz^{\sss N} = - dz^{\sss N} \we dz^{\sss
  M}$. The exterior derivative $d$ acts on the superforms according to
\beq
d \om = \frac{1}{n!} dz^{{\sss M_n}} \we \ldots \we dz^{{\sss M_1}}
\we dz^{\sss N} \pa_{\sss N} \om_{{\sss M_1 \ldots M_n}}.
\eeq
In terms of tangent space indices, these relations become
\beqa
\om & = & \frac{1}{n!} E^{{\sss A_n}} \we \ldots \we E^{{\sss A_1}}
\om_{{\sss A_1 \ldots A_n}} \\
\eqnlab{ext_der}
d \om & = & \frac{1}{n!} E^{{\sss A_n}} \we \ldots \we E^{{\sss A_1}}
\we E^{\sss B} \left( D_{\sss B} \om_{{\sss A_1 \ldots A_n}} +
\frac{n}{2} T_{\sss B A_1}^{\ph{\sss B A_1} {\sss C}} \om_{{\sss C A_2
    \ldots A_n}} \right),
\eeqa
where the torsion two-form $T$ is defined by
\beq
T^{\sss{A}} = dE^{\sss A} = \frac{1}{2} E^{\sss{C}} \we E^{\sss{B}}
T_{{\sss BC}}^{\ph{{\sss BC}}{\sss A}}.
\eeq
Explicitly, we get that the only non-zero component of the torsion is
\beq
T_{\be \ga}^{bc [\al_1 \al_2]} = -2i \de_{\be}^{\ph{\be} [\al_1}
  \de_{\ga}^{\ph{\ga} \al_2]} \Om^{bc}.
\eeq

We will also need to take pull-backs of superforms to bosonic
manifolds embedded in superspace. Let $D$ be such a manifold,
parametrized by $\si^i$,
$i=1,\ldots,\mathrm{dim}(D)$. The pull-back of the differentials in
tangent space is
\beq
E^{\sss A} = d \si^i \pa_i Z^{\sss M} E_{\sss M}^{\sss \ph{M} A}
\equiv d \si^i E_{i}^{\sss A},
\eeq
where $Z^{\sss M}=Z^{\sss M}(\si)$ is the embedding function for the
manifold $D$ in target superspace and $\pa_i$ is the derivative with
respect to $\si^i$. This means that the pull-backs are
\beqa
E^{[\al \be]} & = & d \si^i \left( \pa_i X^{\al
  \be} + i \Om^{ab} \Theta^{[\al}_a \pa_i \Theta^{\be]}_b \right)
\\
E^{\al}_a & = & d \si^i \pa_i \Theta^{\al}_a. 
\eeqa

\section{The interaction terms}

Our goal in this section is to construct supersymmetric interaction
terms that couple a string to a tensor multiplet background. This
background is assumed to be described by the superfield $\Phi^{ab}$
in \Secref{superfields}. In particular, it obeys the constraints
\eqnref{algcon}-\eqnref{diffcon}.

The string world-sheet $\Si$ is embedded in target superspace.
Parametrizing $\Si$ by a set of coordinates $\si^i$, $i=1,2$, the
embedding may be described by a function $Z^{\sss M}(\si)$. 
We denote the components of $Z^{\sss A} = Z^{\sss M} E_{\sss M}^{\sss
  \ph{M} A}$ by $X^{\al \be}$ and
$\Theta^{\al}_a$. Under a supersymmetry transformation
with parameter $\eta$, the embedding functions transform according to
\beqa
\eqnlab{susy_X}
\de X^{\al \be} & = & i \Om^{ab} \eta^{[\al}_a \Theta^{\be]}_b \\
\eqnlab{susy_Th}
\de \Theta_a^{\al} & = & - \eta^{\al}_a.
\eeqa
Note that this means that both $E^{\al \be}$ and $E^{\al}_a$ are
invariant under supersymmetry.

The first coupling term is a standard
Nambu-Goto term, with the string tension replaced by an expression in
the superfield $\Phi^{ab}$, evaluated at the locus of the
string. Explicitly, we have that
\beq
S_{\sss NG} = - \int_{\Si} d^2 \si \sqrt{\Phi \cdot \Phi} \sqrt{- G},
\eeq
where $G$ denotes the determinant of the induced world-sheet metric
$G_{ij}$, given by
\beq
\eqnlab{G_ij}
G_{ij} = \frac{1}{4} \eps_{\al \be \ga \de} E_i^{\al \be} E_j^{\ga
  \de},
\eeq
and the scalar product should be read as
\beq
\Phi \cdot \Phi = \frac{1}{4} \Om_{ac} \Om_{bd} \Phi^{ab} \Phi^{cd}.
\eeq

Under a supersymmetry transformation, $\sqrt{-G}$ is obviously
invariant, while $\sqrt{\Phi \cdot \Phi}$ (being a superfield
evaluated at $Z$) transforms according to
\beq
\de \sqrt{\Phi \cdot \Phi} = \left(
\eta^{\al}_a Q^a_{\al} + \de X^{\al \be} \pa_{\al \be} + \de
\Theta^{\al}_a \pa^a_{\al} \right) \sqrt{\Phi \cdot \Phi} = 0,
\eeq
where we have used \Eqsref{supercharge}, \eqnref{susy_X} and
\eqnref{susy_Th}. Note that this is true only if $\Phi$ obeys the
differential constraint \eqnref{diffcon}. Thus, $S_{\sss NG}$ is
invariant under supersymmetry, modulo terms proportional to the
equations of motion for the free tensor multiplet.

The second coupling term is of Wess-Zumino type and is written as
\beq
S_{\sss WZ} = \int_D F,
\eeq
where the integration is over a three-manifold $D$ such that $\pa D =
\Si$. $F$ is a super-three-form with components
\beqa
F_{[\al_1 \al_2][\be_1 \be_2][\ga_1 \ga_2]} & = & \frac{1}{6} \left(
  H_{\al_1 \be_1} \eps_{\al_2 \be_2 \ga_1 \ga_2} + H_{\be_1 \ga_1}
  \eps_{\be_2 \ga_2 \al_1 \al_2} + H_{\ga_1 \al_1} \eps_{\pa_2 \al_2
    \be_1 \be_2} \right) \\
F^a_{\al [\be_1 \be_2][\ga_1 \ga_2]} & = & \frac{i}{4} \left(
\Psi^a_{\be_1} \eps_{\be_2 \al \ga_1 \ga_2} - \Psi^a_{\ga_1}
\eps_{\ga_2 \al \be_1 \be_2} \right) \\
F^{ab}_{\al \be [\ga_1 \ga_2]} & = & \frac{i}{2} \Phi^{ab} \eps_{\al
  \be \ga_1 \ga_2} \\
F^{abc}_{\al \be \ga} & = & 0,
\eeqa
where antisymmetrization in $\al_1,\al_2$ et.c.~is understood. Using
\Eqnref{ext_der}, it can be shown that
\beqa
dF & = & E^{\de_1 \de_2} \we E^{\ga_1 \ga_2} \we E^{\be_1 \be_2} \we
E^{\al_1 \al_2} \bigg[ \frac{1}{12} \pa_{\al_1 \al_2} H_{\be_1 \ga_1}
\eps_{\be_2 \ga_2 \de_1 \de_2} \bigg] + {} \nn \\
& & {} + E^{\de_1 \de_2} \we E^{\ga_1 \ga_2} \we E^{\be_1 \be_2} \we
E^{\al}_a \bigg[ \frac{i}{4} \eps_{\be_2 \al \de_1 \de_2} \pa_{[\ga_1
      \ga_2} \Psi^a_{\be_1]} + {} \nn \\ & & \qquad {} + \frac{1}{12}
  \eps_{\be_2 \ga_2 \de_1 \de_2} \Big( D^a_{\al} H_{\be_1 \ga_1} - i
  (\pa_{\al \be_1} \Psi^a_{\ga_1} + \pa_{\al \ga_1} \Psi^a_{\be_1} )
  \Big) \bigg] + {} \nn \\ 
& & {} + E^{\de_1 \de_2} \we E^{\ga_1 \ga_2} \we E^{\be}_b \we
E^{\al}_a \bigg[ \frac{i}{4} \eps_{\ga_2 \be \de_1 \de_2} \Big(D^a_{\al}
\Psi^b_{\ga_1} + \Om^{ab} H_{\al \ga_1} - 2 \pa_{\al \ga_1} \Phi^{ab}
\Big) \bigg] + {} \nn \\
& & {} + E^{\de_1 \de_2} \we E^{\ga}_c \we E^{\be}_b \we E^{\al}_a
\bigg[ \frac{i}{4} \eps_{\be \ga \de_1 \de_2} \Big( D^a_{\al}
  \Phi^{bc} + {} \nn \\ & & \qquad {} + i (\Om^{ab}
  \Psi^c_{\al} + \Om^{ca} \Psi^b_{\al} + \frac{1}{2} \Om^{bc}
  \Psi^a_{\al} ) \Big) \bigg] + {} \nn \\ 
& & {} + E^{\de}_d \we E^{\ga}_c \we E^{\be}_b \we E^{\al}_a
\bigg[ \frac{1}{4} \eps_{\al \be \ga \de} \Om^{ab} \Phi^{cd} \bigg].
\eeqa
From \Eqnref{algcon} and \Eqsref{diffcon_Phi}-\eqnref{eom_H} it is
apparent that $F$ is closed (\ie $dF=0$) when the equations of motion
for the superfields are imposed. Invariance under supersymmetry for
the Wess-Zumino term is shown in the same way as for the Nambu-Goto
term.

\section{The local fermionic symmetry}

The fermionic fields $\Theta$ defined above contain twice as many
components as they should (taking supersymmetry into account). The
additional components are removed by the action of a local fermionic
world-sheet symmetry, called $\kappa$-symmetry.

A $\kappa$-transformation of the field $Z^{\sss M}(\si)$ is
\beq
\de_{\kappa} Z^{\sss M} = \kappa^{\sss A} E_{\sss A}^{\sss \ph{A} M},
\eeq
where the (non-constant) parameter $\kappa^{\sss A}$ only contains a
fermionic part $\kappa^{\al}_a$. From this we find the following
$\kappa$-variations of the embedding fields
\beqa
\eqnlab{kappa_X}
\de_{\kappa} X^{\al \be} & = & i \Om^{ab} \kappa^{[\al}_a 
  \Theta^{\be]}_b \\ 
\eqnlab{kappa_Th}
\de_{\kappa} \Theta_a^{\al} & = &  \kappa^{\al}_a.
\eeqa
The corresponding transformation of
the (pull-back of the) tangent space differential $E^{\sss A}$ is
\beq
\de_{\kappa} E^{\sss A} = d \kappa^{\sss A} - \kappa^{\sss B} E^{\sss
  C} T_{\sss C B}^{\sss \ph{C B} A},
\eeq
meaning that
\beqa
\de_{\kappa} E^{[\al \be]} & = & 2i \Om^{ab} \kappa_a^{[\al} d
  \Theta^{\be]}_b \\ 
\de_{\kappa} E^{\al}_a & = & d \kappa^{\al}_a.
\eeqa
In these expressions, the exterior derivative $d$ should be understood
as $d=d\si^i \pa_i$.

The parameters $\kappa$ are subject to the constraint
\beq
\eqnlab{kappa_cond}
\Ga^{\al}_{\ph{\al} \be} \kappa^{\beta}_a = \ga_a^{\ph{a} b}
\kappa^{\al}_b,
\eeq
where
\beqa
\Ga^{\al}_{\ph{\al} \be} & = & \frac{1}{2} \frac{1}{\sqrt{- G}}
\eps^{ij} E_i^{\al \ga} E_j^{\de \eps} \eps_{\be \ga \de \eps} \\
\ga_a^{\ph{a} b} & = & \frac{1}{\sqrt{\Phi \cdot \Phi}} \Om_{ac}
\Phi^{cb}.
\eeqa
Obviously, $\Ga^{\al}_{\ph{\al} \al}=0$ and $\ga_a^{\ph{a} a}=0$. One
may also show that $\Ga^{\al}_{\ph{\al} \be} \Ga^{\be}_{\ph{\be} \ga}
= \de^{\al}_{\ph{\al} \ga}$ and $\ga_a^{\ph{a} b} \ga_b^{\ph{b} c} =
\de_a^{\ph{a} c}$. This means that the condition \eqnref{kappa_cond}
eliminates half of the components in $\kappa$.

The terms $S_{\sss NG}$ and $S_{\sss WZ}$ are not $\kappa$-symmetric
by themselves, but a certain linear combination of them is. This is
possible if the $\kappa$-variation of $F$ is not only closed, but also
exact, \ie
\beq
\de_{\kappa} F = d \omega_{\kappa}.
\eeq
By Stokes' theorem, the variation of the Wess-Zumino term becomes
\beq
\de_{\kappa} S_{\sss WZ} = \int_{\Si} \omega_{\kappa}.
\eeq
Explicitly, we find that
\beqa
\de_{\kappa} F & = & \frac{1}{3!} \Big( 3 d(\de_{\kappa} Z^{\sss P})
\we dZ^{\sss N} \we dZ^{\sss M} F_{\sss MNP} + dZ^{\sss P} \we
dZ^{\sss N} \we dZ^{\sss M} \de_{\kappa} Z^{\sss Q} \pa_{\sss Q}
F_{\sss MNP} \Big) \nn \\
& = & \frac{1}{2} d \left( \de_{\kappa} Z^{\sss P} dZ^{\sss N} \we
dZ^{\sss M}  F_{\sss MNP} \right),
\eeqa
where we have used that $dF=0$, which is valid when the tensor
multiplet fields obey the free equations of motion. This means that
\beq
\om_{\kappa} = d\si^{i} \we d\si^j \eps_{\al \be \ga \de}
\kappa^{\al}_a E^{\be \ga}_i \left( \frac{i}{4} E_j^{\de \veps}
\Psi^a_{\veps} - \frac{i}{2} \pa_j \Theta^{\de}_b \Phi^{ab} \right).
\eeq

We now turn to the variation of the Nambu-Goto term. We find, using
\Eqnref{kappa_cond}, that
\beqa
\de_{\kappa} \sqrt{\Phi \cdot \Phi} & = & \frac{i}{4} \frac{1}{\sqrt{-
    \det G}} \eps^{ij} \eps_{\al \be \ga \de} \kappa^{\al}_a E^{\be
  \ga}_i E_j^{\de \veps} \Psi^a_{\veps} \\
\de_{\kappa} \sqrt{- \det G} & = & - \frac{i}{2} \frac{1}{\sqrt{\Phi
    \cdot \Phi}} \eps^{ij} \eps_{\al \be \ga \de} \kappa^{\al}_a E^{\be
  \ga}_i \pa_j \Theta^{\de}_b \Phi^{ab} .
\eeqa
This yields immediately that
\beq
\de_{\kappa} \left( S_{\sss NG} + S_{\sss WZ} \right) = 0.
\eeq
Thus, this specific linear combination of coupling terms is both
$\kappa$-symmetric and supersymmetric. This means that the interaction
\beq
S = - \int_{\Si} d^2 \si \sqrt{\Phi \cdot \Phi} \sqrt{- G} + \int_D F
\eeq
describes the supersymmetric coupling of a self-dual string to a
tensor multiplet background.

\vspace{1cm}
\noindent
\textbf{Acknowledgments:} M.H.~is a Research Fellow at the Royal
Swedish Academy of Sciences.

We would like to thank Martin Cederwall for many helpful comments,
and regret that we did not follow more of his advice earlier.

\clearpage
\bibliographystyle{utphysmod3b}
\bibliography{coupled}

\end{document}